\documentclass[a4paper,twoside,11pt]{article} \usepackage[journal]{optional}
\usepackage {amsmath,amssymb,latexsym}
\usepackage {delarray,graphics,graphpap,alltt,graphicx}
\newcommand{\qedllncs}{\opt{cwa,cna}{\qed}}

\newcommand{\FST}{\mathrm{FST}}

\newcommand{\muE}[1]{\mu \left( \left. #1 \right| \E \right)}

\newcommand{\proj}{\mathrm{proj}}
\newcommand{\fst}{\mathrm{FST}}
\newcommand{\fsd}[1]{\mathrm{D^{\mathnormal{#1}}_{FS}}}

\title{{\bf Feasible Depth}}

\opt{journal}{
    \newcommand{\email}[3]{{\tt #1} (at) {\tt #2} (dot) {\tt #3}}
    \usepackage{a4wide}
    \usepackage{commands}
    \usepackage{numbering}
    \author{
        David Doty\thanks{Department of Computer Science, Iowa State University, Ames, IA 50011 USA. ddoty (at) iastate (dot) edu. This author was partially supported by grant number 9972653 from the National Science Foundation as part of their Integrative Graduate Education and Research Traineeship (IGERT) program.}
        \and
        Philippe Moser\thanks{Dept de Inform\'atica e Ingenier\'{\i}a de Sistemas, Centro Polit\'ecnico Superior, Zaragoza, Spain. mosersan (at) gmail (dot) com. This author was partially supported by subvenciones para grupos de investigaci\'on Gobierno de Arag\'on UZ-T27 and subvenciones de fomento de movilidad Gobierno de Arag\'on MI31/2005.}
    }
}

\opt{cwa,cna}{
    \usepackage{commands}
    \numberwithin{equation}{section}
    \numberwithin{theorem}{section}
    \spnewtheorem{cor}[theorem]{Corollary}{\bfseries}{\itshape}
    \spnewtheorem{lem}[theorem]{Lemma}{\bfseries}{\itshape}
    \spnewtheorem{obs}[theorem]{Observation}{\bfseries}{\itshape}
    \spnewtheorem{prop}[theorem]{Proposition}{\bfseries}{\itshape}
    \spnewtheorem{defn}[theorem]{Definition}{\bfseries}{\itshape}
    \spnewtheorem*{rem}{Remark}{\bfseries}{\itshape}
    \spnewtheorem*{ack}{Acknowledgment}{\bfseries}{\rmfamily}
    \spnewtheorem*{acks}{Acknowledgments}{\bfseries}{\rmfamily}

    \author{
        {David Doty\inst{1}\thanks{Corresponding author. This author was partially supported by grant number 9972653 from the National Science Foundation as part of their Integrative Graduate Education and Research Traineeship (IGERT) program.}}
        \and
        {Philippe Moser\inst{2}\thanks{This author was partially supported by subvenciones para grupos de investigaci\'on Gobierno de Arag\'on UZ-T27 and subvenciones de fomento de movilidad Gobierno de Arag\'on MI31/2005.}}
    }
    \institute{ Department of Computer Science, Iowa State University, Ames, IA 50011, USA. \email{ddoty (at) iastate (dot) edu}
    \and
    Dept de Inform\'atica e Ingenier\'{\i}a de Sistemas, Centro Polit\'ecnico Superior, Zaragoza, Spain. \email{mosersan (at) gmail (dot) com}}
}

\date{}

\begin{document}
    \maketitle


\begin{abstract}
    This paper introduces two complexity-theoretic formulations of Bennett's logical depth: \emph{finite-state depth} and \emph{polynomial-time depth}.
    It is shown that for both formulations, trivial and random infinite sequences are shallow, and a \emph{slow growth law} holds, implying that deep sequences cannot be created easily from shallow sequences.
    Furthermore, the $\E$ analogue of the halting language is shown to be polynomial-time deep, by proving a more general result: every language to which a nonnegligible subset of $\E$ can be reduced in uniform exponential time is polynomial-time deep.
\end{abstract}

\opt{cwa,cna}{\noindent {\bf Keywords}: dimension, depth, randomness, polynomial-time, finite-state}

\section{Introduction}
Whereas many structures found in nature are highly complex (a DNA sequence, a cell), some seem much simpler, either because of their complete regularity (ice), or their complete randomness (gas). Bennett introduced logical depth \cite{Benn88} to formalize computationally the difference between complex and non-complex (trivial or random) structures. Briefly, a logically deep object is one with a shorter description than itself, but which requires a long time to compute from this short description.

Depth is not a measure of information contained in an object, which correlates with \emph{randomness}, but rather its value, or its \emph{useful} information content. According to classical \cite{Shannon48} or algorithmic information theory \cite{LiVi97}, the information content of a sequence is not representative of its value. Consider an infinite binary sequence produced by random coin tosses. Although the sequence contains a large amount of information in the sense that, with probability 1, it cannot be significantly compressed, its information is not of much value, except as a source of input to randomized algorithms. Contrast this with the characteristic sequence of the halting language, access to which enables any computably enumerable language to be decided in linear time. From this perspective, the halting sequence is much more useful than a randomly generated sequence.

Bennett's logical depth separates the sequences that are deep (i.e., that show high internal organization) from those that are shallow (i.e., not deep). Informally, deep sequences are those which contain redundancy, but in such a way that an algorithm requires extensive resources to exploit the redundancy (for instance, to compress or to predict the sequence). In other words, deep sequences are organized, but in a nontrivial way. Highly redundant sequences like 00000... are shallow, because they are trivially organized. Random sequences are shallow, because they are completely unorganized. One of the key features of Bennett's logical depth is that it obeys a \emph{slow growth law} \cite{Benn88,Lutz:CDR}: no fast process can transform a shallow sequence into a deep one. Therefore a deep object can be created only through a complex, time-consuming process.

Bennett \cite{Benn88} showed that the halting language is deep, arguing that its depth was evidence of its usefulness.
Juedes, Lathrop, and Lutz \cite{Lutz:CDR} generalized this result and solidified the connection between usefulness and depth by proving that every \emph{weakly useful} language \cite{Lutz:WUS} is deep, where a weakly useful language is one to which a nonnegligible subset of the decidable languages (in the sense of resource-bounded measure theory \cite{Lutz:AEHNC}) reduce in a fixed computable time bound.

Unfortunately, because it is based on Kolmogorov complexity, Bennett's logical depth is not computable.
Lathrop and Lutz \cite{Lutz:RCD} investigated \emph{recursive computational depth}, which is computable, but not within any feasible time scale.
Antunes, Fortnow, van Melkebeek, and Vinodchandran \cite{AFVV06} investigated several polynomial-time formulations of depth as instances of the more general concept of computational depth obtained by considering the difference between variants of Kolmogorov complexity. Deep and intriguing connections were demonstrated between depth and average-case complexity, nonuniform circuit complexity, and efficient search for satisfying assignments to Boolean formulas. Nevertheless, some of the depth notions in \cite{AFVV06} require complexity assumptions to prove the existence of deep sequences, and not all the depth notions obey slow growth laws. Furthermore, \cite{AFVV06} lacks a polynomial-time analogue of the Juedes-Lathrop-Lutz theorem demonstrating that useful objects are necessarily deep.

The aim of this paper is to propose a feasible depth notion that satisfies a slow growth law and in which deep sequences can be proven to exist. We propose two such notions: finite-state depth, and polynomial-time depth. Furthermore, we connect polynomial-time depth to usefulness in deciding languages in the complexity class $\E$. In both cases, the definition of depth intuitively reflects that of Bennett's logical depth: a sequence is deep if it is redundant, but an algorithm requires extensive resources in order to exploit the redundancy.

Our formulation of finite-state depth is based on the classical model of finite-state compressors and decompressors introduced by Shannon \cite{Shannon48} and investigated by Huffman \cite{Huff59a} and Ziv and Lempel \cite{ZivLem78}. Informally, a sequence is finite-state deep if given more states, a finite-state machine can decompress the sequence from an input significantly shorter than is possible with fewer states. We show that both finite-state trivial sequences (sequences with finite-state strong dimension \cite{Athreya:ESDAICC} equal to 0) and finite-state random sequences (those with finite-state dimension \cite{Dai:FSD} equal to 1, or equivalently normal sequences \cite{Bore09}) are shallow. Our main result in this section shows that finite-state depth obeys a slow growth law: no information lossless finite-state transducer can transform a finite-state shallow sequence into a finite-state deep sequence. We conclude the section by proving the existence of finite-state deep sequences.

Our formulation of polynomial-time depth -- contrary to finite-state depth -- is not based on compression algorithms but on polynomial-time oblivious predictors.
Given a language $L$, a polynomial-time oblivious predictor is a polynomial-time computable function that, given an input string $x$, predicts the probability that $x \in L$. Informally, $L$ is polynomial-time deep if, given more time, a predictor is better able to predict membership of strings in $L$. We show that both $\E$-trivial languages (languages in the complexity class $\E$) and $\E$-random languages are polynomial-time shallow. Our main results in this section are a slow growth law similar to that for finite-state depth and logical depth, and a theorem stating that any language which is ``useful'' for quickly deciding languages in $\E$ must be polynomial-time deep. It follows that $H_\E$, the $\E$ version of the halting language, is polynomial-time deep.

\section{Preliminaries}
    \opt{cwa}{Proofs of some results are found in the Appendix.}

    $\N$ is the set of all nonnegative integers. A \emph{(finite) string} is an element of $\binary$. An \emph{(infinite) sequence} is an element of the Cantor space $\C = \binary[\infty]$. For a string or sequence $S$ and $i,j\in\N$, $S[i \twodots j]$ denotes the substring consisting of the $\ith$ through the $j^\mathrm{th}$ bits of $S$, inclusive, and $S \pre n$ denotes $S[0 \twodots n-1]$. For a string $x$ and a string or sequence $S$, we write $x \prefix S$ to denote that $x = S \pre n$ for some $n\in\N$. For a string $x$, its length is denoted by $|x|$. $s_0 ,s_1 , s_2 \ldots $ denotes the standard enumeration of the strings in $\bool{*}$ in lexicographical order, where $s_0 = \lambda$ denotes the empty string. If $x,y$ are strings, we write $x < y$ if $|x|<|y|$ or $|x|=|y|$ and $x$ precedes $y$ in alphabetical order, and $x \leq y$ if $x < y$ or $x=y$.

    A \emph{language} is a subset of $\binary$.
    A \emph{class} is a set of languages.
    The \emph{characteristic sequence} of a language $L$ is the sequence $\chi_L \in\bool{\infty}$, whose $\nth$ bit is 1 if and only if $s_n \in L$. Because $L \mapsto \chi_L$ is a bijection, we will often speak of languages and sequences interchangeably, with it understood that the ``sequence'' $L$ refers to $\chi_L$, and the ``language'' $\chi_L$ refers to $L$.
    \opt{journal,cwa}{For $n\in\N,$ we write $L \pre n$ to denote $\chi_L \pre n$. Given $s_n\in\binary$, let $L(s_n)=\chi_L[n]$ (the value 1 if $s_n \in L$, and 0 if $s_n \not\in L$).}
    Let $\E = \bigcup_{c\in\N}\DTIME(2^{cn})$ and $\EXP = \bigcup_{c\in\N}\DTIME(2^{n^c})$.

    Let $1 \leq i \leq j \in \N$. The $\ith$ projection function $\proj_i : (\bool{*})^j \rightarrow \bool{*}$, is given by
    $\proj_i(x_1,\ldots,x_j) = x_i$.

\section{Finite-State Depth}

\subsection{Finite-State Compression}

We use a model of finite-state compressors and decompressors based on finite-state transducers, which was introduced in a similar form by Shannon \cite{Shannon48} and investigated by Huffman \cite{Huff59a} and Ziv and Lempel \cite{ZivLem78}. Kohavi \cite{Koha78} gives an extensive treatment of the subject.

A \emph{finite-state transducer (FST)} is a 4-tuple $T = (Q,\delta,\nu,q_0),$ where
\begin{itemize}
    \item $Q$ is a nonempty, finite set of \emph{states},

    \item $\delta: Q \times \bool{} \to Q$ is the \emph{transition function},

    \item $\nu: Q \times \bool{} \to \bool{*}$ is the \emph{output function},

    \item $q_0 \in Q$ is the \emph{initial state}.
\end{itemize}
Furthermore, we assume that every state in $Q$ is reachable from $q_0$.

For all $x\in\bool{*}$ and $a\in\bool{}$, define the \emph{extended transition function} $\widehat{\delta}:\bool{*} \to Q$ by the recursion
    $\widehat{\delta}(\lambda) = q_0$, and $\widehat{\delta}(xa) = \delta(\widehat{\delta}(x),a).$
For $x\in\bool{*}$, we define the \emph{output} of $T$ on $x$ to be the string $T(x)$ defined by the recursion
    $T(\lambda) = \lambda$, and $T(xa) = T(x)\nu(\widehat{\delta}(x),a)$
for all $x\in\bool{*}$ and $a\in\bool{}$.

A FST can trivially act as an ``optimal compressor'' by outputting $\lambda$ on every transition arrow, but this is, of course, a useless compressor, because the input cannot be recovered. A FST $T = (Q,\delta,\nu,q_0)$ is \emph{information lossless (IL)} if the function $x \mapsto (T(x),\widehat{\delta}(x))$ is one-to-one; i.e., if the output and final state of $T$ on input $x\in\binary$ uniquely identify $x$. An \emph{information lossless finite-state transducer (ILFST)} is a FST that is IL. We write FST to denote the set of all finite-state transducers, and we write ILFST to denote the set of all information lossless finite-state transducers.
We say $f:\binary[\infty]\to\binary[\infty]$ is \emph{FS computable} (resp. \emph{ILFS computable}) if there is a FST (resp. ILFST) $T$ such that, for all $S\in\binary[\infty]$, $\limn |T(S \pre n)| = \infty$ and, for all $n\in\N$, $T(S \pre n) \prefix f(S)$. In this case, define $T(S) = f(S)$.

The following well-known theorem \cite{Huff59a,Koha78} states that the function from $\bool{*}$ to $\bool{*}$ computed by an ILFST can be inverted -- in an approximate sense -- by another ILFST.

\begin{theorem} \label{thm.il.reverse}
For any ILFST $T$, there exists an ILFST $T^{-1}$ and a constant $c\in\N$ such that, for all $x\in\bool{*}$,
$x \harp (|x|-c) \sqsubseteq T^{-1}(T(x)) \sqsubseteq x$.
\end{theorem}

\begin{cor}
For any ILFST $T$, there exists an ILFST $T^{-1}$ such that, for all sequences $S$, $T^{-1}(T(S)) = S$.
\end{cor}

Fix some standard binary representation $\sigma_T\in\bool{*}$ of each FST $T$, and define $|T| = |\sigma_T|$. For all $k\in\N$, define
$$
\begin{array}{lcl}
    \mathrm{FST}^{\leq k} &=& \{ T \in \mathrm{FST}:|T| \leq k\},\\
    \mathrm{ILFST}^{\leq k} &=& \{ T \in \mathrm{ILFST}:|T| \leq k\}.
\end{array}
$$

Let $k\in\N$ and $x\in\bool{*}$. The \emph{$k$-FS decompression complexity} (or when $k$ is clear from context, \emph{FS complexity}) of $x$ is
$$
    \fsd{k}(x) = \min_{p\in\bool{*}} \setr{|p|}{(\exists T \in \FST^{\leq k})\ T(p) = x},
$$
i.e., the size of the smallest program $p\in\bool{*}$ such that some $k$-bit FST outputs $x$ on input $p$.

For a fixed $k$, $\fsd{k}$ is a finite state analogue of Kolmogorov complexity. For any sequence $S$, define the \emph{finite-state dimension} of $S$ by
\begin{equation}\label{e.equivalence}
    \dimfs(S)
    =\lim_{k\rightarrow\infty} \liminf_{n\rightarrow\infty} \frac{\fsd{k}(S\harp n)}{n},
\end{equation}
and the \emph{finite-state strong dimension} of $S$ by
\begin{equation}\label{e.equivalence.str}
    \Dimfs(S)
    =\lim_{k\rightarrow\infty} \limsup_{n\rightarrow\infty} \frac{\fsd{k}(S\harp n)}{n}.
\end{equation}

Finite-state dimension and strong dimension measure the degree of finite-state randomness of a sequence. The above definitions are equivalent \cite{SheLemZiv95,DotMos06} to several other definitions of finite-state dimension and strong dimension in terms of finite-state gamblers \cite{Dai:FSD,Athreya:ESDAICC}, entropy rates \cite{ZivLem78,Bourke:ERFSD}, information lossless finite-state compressors \cite{ZivLem78,Dai:FSD,Athreya:ESDAICC}, and finite-state log-loss predictors \cite{Hitchcock:FDLLU}.

Schnorr and Stimm \cite{SchSti72} (and more explicitly, Bourke, Hitchcock, and Vinodchandran \cite{Bourke:ERFSD}) showed that a sequence has finite-state dimension 1 if and only if it is \emph{normal} in the sense of Borel \cite{Bore09}, meaning that for all $k\in\N$, every substring of length $k$ occurs in $S$ with limiting frequency $2^{-k}$.

\newcommand{\lemmaOneAndTwo}{The next two lemmas show that ILFST's cannot alter the FS complexity of a string by very much.

\begin{lem}\label{lem-l.1} Let $M$ be an ILFST. Then
$$
    (\exists c_1\in\N) (\forall k\in\N) (\forall x\in\bool{*})\ \fsd{k+c_1}(M(x))\leq \fsd{k}(x).
$$
\end{lem}

\begin{proof}
    The proof idea of the lemma is the following. Let $k,x$ be in the statement of the lemma, let $p$ be a $k$-minimal program for $x$, i.e.
    $A(p)=x$ where $A\in\fst^{\leq k}$, and $\fsd{k}(x)=|p|$. We construct $A'$ and $p'$ for $M(x)$.
    Let $p'=p$ and let $A'$ be the automata which on input $p'$ simulates $A(p)$, and plugs the output
    into $M$. The size of $A'$ is roughly the size of $A$ plus the size of $M$, i.e. $\fsd{k+c_1}(M(x))\leq \fsd{k}(x)$,
    for some constant $c_1$.
    More formally, let
    \[
        \delta_A:Q_A \times \bool{} \rightarrow Q_A
    \]
        be the transition function of $A$,
    with
    \[
        Q_C=\{(q_i,s_i)| \ 1\leq i \leq t_C \}\subset (\mathcal{P}(\bool{*}))^2 \qquad C\in\{A,M\}
    \]
    where $q_i\in\bool{*}$ are the states and $s_i\in\bool{*}$ are the corresponding output strings,
    and let $\delta_M:Q_M\times\bool{}\rightarrow Q_M$ be the transition function  for $M$.
    We construct $\delta':Q' \times \bool{} \rightarrow Q'$ for $A'$.
    Let
    \[
        Q'=Q_A\times Q_M \times \{A,M\}\times\bool{\leq t}\times\bool{\leq t}
    \]
    where $t$ is a constant depending on $A$ and $M$.
    Let $(q_A,s_A)\in Q_A$, $(q_M,s_M)\in Q_M$, $s,m\in\bool{\leq t}$ and $b\in\bool{}$.
    Define
    \[
    \begin{cases}
        \delta'((q_A,s_A),(q_M,s_M),A,s,m,b) \\ \ \ \ \ = (\delta_A((q_A,s_A),b),(q_M,s_M),M,\lambda,\proj_2(\delta_A((q_A,s_A),b)))\\
        \delta'((q_A,s_A),(q_M,s_M),M,s,m,b) \\  \ \ \ \ = ((q_A,s_A),\delta_M((q_M,s_M),m),A,\proj_2(\delta_M((q_M,s_M),m)),\lambda) .
    \end{cases}
    \]
\qedllncs\end{proof}

\begin{lem}\label{lem-l.2} Let $M$ be an ILFST. Then
$$
    (\exists c_2\in\N) (\forall k\in\N) (\forall x\in\bool{*})\ \fsd{k+c_2}(x)\leq \fsd{k}(M(x)).
$$
\end{lem}

\begin{proof}
    The proof is similar to Lemma \ref{lem-l.1}. Let $k,x$ be as in the statement of the lemma. By Theorem \ref{thm.il.reverse}, there exists an ILFST $M^{-1}$ and a constant $b$ such that for any string $x$, $x\harp |x|-b \sqsubseteq M^{-1}(M(x)) \sqsubseteq x$.

    Let $p$ be a $k$-minimal program for $M(x)$, i.e.
    $A(p)=M(x)$ where $A\in\fst^{\leq k}$, and $\fsd{k}(M(x))=|p|$. We construct $A'$ and $p'$ for $x$.
    Let $y=M^{-1}(M(x))$, i.e. $yz=x$ and $|z|\leq b$.
    Let $p'=p$ and let $A'$ be the automata which on input $p'$ simulates $A(p)$, plugs the output
    into $M^{-1}$ and adds  $z$ at the end of $M^{-1}$'s output.
    The size of $A'$ is roughly the size of $A$ plus the size of $M$ plus the size of $z$
    (which is of size at most $b$), i.e. $\fsd{k+c_2}(M(x))\leq \fsd{k}(x)$, for some constant $c_2$.
\qedllncs\end{proof}
}

\opt{journal}{\lemmaOneAndTwo}

\subsection{Finite-State Depth}
    Intuitively, a sequence is finite-state deep if a finite state transducer, given additional states (or more accurately, additional bits with which to represent the transducer), can decompress the sequence from a significantly shorter input.

    \begin{defn}
    A sequence $S$ is \emph{finite-state deep} if
        \[
            (\exists \alpha>0) (\forall k\in\N) (\exists k'\in\N) (\existsio n\in\N)\ \fsd{k}(S \pre n) - \fsd{k'}(S \pre n) \geq \alpha n.
        \]
    A sequence $S$ is \emph{finite-state shallow} if it is not finite-state deep.
    \end{defn}
    \begin{rem}
    All results in this section remain true if the quantification in the definition of finite-state depth is changed to $$(\forall k\in\N) (\exists \alpha>0) (\exists k'\in\N) (\existsio n\in\N)\ \fsd{k}(S \pre n) - \fsd{k'}(S \pre n) \geq \alpha n.$$ Note that any sequence deep by the former definition must be deep by the latter definition.
    \end{rem}

    Finite-state trivial and finite-state random sequences are finite-state shallow.

    \begin{prop} \label{propFSTrivRandShallow}
        Let $S\in\C$.
        \begin{enumerate}
        \item If $\Dimfs(S)=0$, then $S$ is finite-state shallow.
        \item If $S$ is normal (i.e., if $\dimfs(S)=1$), then $S$ is finite-state shallow.
        \end{enumerate}
    \end{prop}

\newcommand{\proofPropFSTrivRandShallow}{
    Let $S\in\C$ satisfy $\Dimfs(S)=0$ and let $\alpha>0$. By (\ref{e.equivalence.str}) let $k\in\N$ be such that
    \[
        \limsup_{n\rightarrow\infty} \frac{\fsd{k}(S \pre n)}{n} < \alpha,
    \]
    i.e. $(\forallio n\in\N)$ $\fsd{k}(S \pre n) < \alpha n$. Therefore
    \[
        (\forall k'\in\N)(\forallio n\in\N) \ \fsd{k}(S \pre n) - \fsd{k'}(S \pre n)
        \leq \fsd{k}(S \pre n) < \alpha n.
    \]
    Since $\alpha$ is arbitrary, $S$ is finite-state shallow.

    Let $S\in\C$ be normal, $k\in\N$, and $\alpha>0$. Because normal sequences have finite-state dimension 1,
    \[
        (\forall k'\in\N)(\forallio n\in \N)\ \fsd{k'}(S \pre n)> \left( 1-\alpha \right) n.
    \]
    Thus
    \[
        (\forall k'\in\N)(\forallio n \in\N) \ \fsd{k}(S \pre n) - \fsd{k'}(S \pre n) < n - \left( 1-\alpha \right) n  = \alpha n .
    \]
    Because $\alpha$ is arbitrary, $S$ is finite-state shallow.
}

\opt{journal}{\begin{proof}\proofPropFSTrivRandShallow\end{proof}}

    Finite-state deep sequences cannot be created easily, as the following theorem shows. More precisely, no ILFST can transform a finite-state shallow sequence into a finite-state deep sequence.

    \begin{theorem}[Finite-state slow growth law] \label{thm-sgl}
        Let $S$ be any sequence, let $f:\bool{\infty}\rightarrow \bool{\infty}$ be ILFS computable, and let $S'=f(S)$. If $S'$ is finite-state deep, then $S$ is finite-state deep.
    \end{theorem}

\newcommand{\proofThmSGL}{
        Let $S,S',f$ be as in the statement of the lemma and $M$ be an ILFST computing $f$. Because $S'$ is finite-state deep,
        \begin{equation}\label{e.depth1}
            (\exists \alpha > 0) (\forall k\in\N) (\exists k'\in\N) (\existsio n\in \N)\ \fsd{k}(S' \pre n) -\fsd{k'}(S' \pre n)\geq \alpha n.
        \end{equation}
        Let $l\in\N$ and let $c=\max\{c_1,c_2\}$ where $c_1,c_2$ are the two constants in Lemmas \ref{lem-l.1} and \ref{lem-l.2}. Let $l'=k'+c$ where $k'$ is obtained from (\ref{e.depth1}) with $k=l+c$. For all $n\in\N$, denote by $m_n$ the smallest integer such that $M(S\harp m_n) = S'\harp n$. Because $M$ is IL, it cannot visit a state twice without outputting at least one bit, so there exists a constant $\beta > 0$ such that, for all $n\in\N$, $n \geq \beta m_n$.
        For infinitely many $n\in\N$,
        \begin{align*}
            &\ \ \ \  \fsd{l}(S \pre m_n) -\fsd{l'}(S \pre m_n) \\
            &= \fsd{l}(S \pre m_n) -\fsd{k'+c}(S \pre m_n) \qquad &l'=k'+c\\
            &\geq \fsd{l}(S \pre m_n) -\fsd{k'}(M(S \pre m_n)) \qquad &\text{Lemma \ref{lem-l.2}}\\
            &= \fsd{k-c}(S \pre m_n) -\fsd{k'}(M(S \pre m_n)) \qquad &k=l+c\\
            &\geq \fsd{k}(M(S \pre m_n)) -\fsd{k'}(M(S \pre m_n)) \qquad &\text{Lemma \ref{lem-l.1}}\\
            &= \fsd{k}(S' \pre n)) -\fsd{k'}(S' \pre n)) \qquad &\text{definition of $m_n$}\\
            &\geq \alpha n \qquad &\text{by (\ref{e.depth1})}\\
            &\geq \alpha \beta m_n,  \qquad &\text{because $M$ is IL}
        \end{align*}
        whence $S$ is finite-state deep.
}
\opt{journal}{\begin{proof}\proofThmSGL\end{proof}}

\newcommand{\thmExistsFSDSetup}{
    We next prove the existence of finite-state deep sequences. We require two technical lemmas first, which place bounds on the FS complexity of two concatenated strings.

    \begin{lem}\label{l.4}
        $(\forall l\in\N) (\forall x,y\in\bool{*})\
            \fsd{l}(xy)\geq \fsd{l}(x) + \fsd{l}(y) - 2^l.$
    \end{lem}

    \begin{proof}
    Let $l,x,y$ be as in the statement of the lemma and suppose $\fsd{l}(xy)= |pp'|$ where $T\in\FST^{\leq l}$, $p,p'\in\bool{*}$, with $T(pp')=xy$, and $T(p)=x$.
    Thus $\fsd{l}(x)\leq |p|$ and there exists $s\in\bool{\leq 2^l}$ (because $T$ has less than $2^l$ states) such that
    $T(sp')=y$, i.e. $\fsd{l}(y)\leq |p'|+2^l$. Therefore
    \[
        \fsd{l}(xy) = |p| + |p'| \geq \fsd{l}(x) + \fsd{l}(y) - 2^l,
    \]
    which proves the lemma.
    \qedllncs\end{proof}

    \begin{lem}\label{l.5}
        $(\exists c\in\N) (\forall l\in\N) (\forall x,y\in\bool{*})\
            \fsd{l+c}(xy)\leq 2|x| + \fsd{l}(y) +2.$
    \end{lem}

    \begin{proof}
    Let $l,x,y$ be as in the statement of the lemma and let $p$ be a minimal program for $y$, i.e.
    $\fsd{l}(y)= |p|$ where $A(p)=y$ with $p\in\bool{*}$ and $A\in \fst^{\leq l}$.
    Let $p'=x'01p$ where $x'$ is $x$ with every bit doubled
    and let $A'\in\fst^{\leq l+c}$ where $c$ is a constant independent of $l$ be the following FST for $xy$:
    $A(p')$ uses $d(x)$ to output $x$, then upon reading $01$,
    it outputs $A(p)$.
    \qedllncs\end{proof}
}

\opt{journal}{\thmExistsFSDSetup}
\opt{cwa}{Finally, we show that finite-state deep sequences exist.}

    \begin{theorem} \label{thm-exists-fsd}
        There exists a finite-state deep sequence.
    \end{theorem}

\newcommand{\proofThmExistsFSD}{
    For all $r\in\binary$, define the FST $T_r=(\{ q_0 \} , \delta, \nu, q_0)$, where, for $b\in\{0,1\}$, $\delta(q_0,b) = q_0$ and $\nu(q_0,b) = r$. Define the constant $c' = |T_r| - |r|$ (i.e., the number of extra bits beyond $r$ required to describe $T_r$; note that this is a constant independent of $r$).

    We construct the finite-state deep sequence $S=S_1S_2\ldots$ in stages, with $S_i\in\binary$ for all $i\in\N$.
    Let $\phi:\N \rightarrow \N\times\N$ be a function such that $(\forall k \in\N) (\existsio j \in\N)$
    $\phi(j)=(k,2^{2^{k+1}})$, and for all $i\in\N$, $\proj_2(\phi(i))=2^{2^{\proj_1 (\phi(i))+1}}$.
    Let $j\in\N$ and suppose the prefix $S_1S_2\ldots S_{j-1}$ has already been constructed.
    Let $t_{j-1}=|S_1S_2\ldots S_{j-1}|$.
    Let $(k,k')=\phi(j)$, so that $k'=2^{2^{k+1}}$.

    Intuitively, at stage $j$, we will diagonalize against $k$-bit FST's to make $\fsd{k}(S_1 \ldots S_j)$ large, while helping a particular $(k'+c)$-bit FST ($c$ the constant from Lemma \ref{l.5}) so that $\fsd{k'+c}(S_1 \ldots S_j)$ is small.

    Let $r_j\in \bool{k' - c'}$ be $k$-FS-random in the sense that
    \begin{equation}\label{e.defder}
        \fsd{k} (r_j) \geq |r_j|-2^{k/2} .
    \end{equation}
    Note that such a string always exists because there are at most
    $|\fst^{\leq k}| \cdot 2^{|r_j|-2^{k/2}}<2^{|r_j|}$ strings contradicting \eqref{e.defder}.
    Let $u_j = 12 t_{j-1}$.
    Let $S_j=r_j^{u_j / |r_j|}$ be $u_j / |r_j|$ consecutive copies of $r_j$. Let $T=T_{r_j}$ as described above. Then $|T| = k'$. It is clear that $T$ outputs $S_j = r_j^{u_j/|r_j|}$ on any input program of length $u_j/|r_j|$. Therefore $\fsd{k'}(S_j) \leq u_j/|r_j|.$
    Lemma \ref{l.5} implies that
    $$
    \fsd{k'+c}(S_1\ldots S_j) \leq 2 |S_1\ldots S_{j-1}| + \fsd{k'}(S_j) + 2,
    $$
    whence
    \begin{equation} \label{eq-1-thm-fsdeep-exist}
        \fsd{k'+c}(S_1\ldots S_j) \leq 2 t_{j-1} + \frac{u_j}{|r_j|} + 2.
    \end{equation}
    Note that
    \begin{align}
        \fsd{k}(S_1\ldots S_j) &\geq \fsd{k} (S_1\ldots S_{j-1}) + \fsd{k}(S_j) - 2^k \qquad &\text{Lemma \ref{l.4}}
    \nonumber \\ &\geq
        \fsd{k}(S_j) - 2^k
    \nonumber \\ &\geq
        \frac{u_j}{|r_j|} \fsd{k}(r_j) - \left(\frac{u_j}{|r_j|} + 1 \right) 2^k \qquad &\text{Lemma \ref{l.4}}
    \nonumber \\ &\geq
        u_j - \frac{u_j}{|r_j|} 2^{k/2} - \left(\frac{u_j}{|r_j|} + 1 \right) 2^k  \qquad &\text{choice of $r_j$}
    \nonumber \\ &\geq
        u_j - \frac{u_j}{|r_j|} 2^{k+1}
    \nonumber \\ &>
        u_j \left( 1 - \frac{2^{k+1}}{k'} \right).         \label{eq-2-thm-fsdeep-exist}
    \end{align}
    By \eqref{eq-1-thm-fsdeep-exist} and \eqref{eq-2-thm-fsdeep-exist},
    \begin{align*}
        &\ \ \ \ \fsd{k}(S_1\ldots S_j) - \fsd{k'+c}(S_1\ldots S_j)
        \\&\geq
        u_j \left(1 - \frac{2^{k+1}}{k'} - \frac{1}{|r_j|} \right) - 2 t_{j-1} - 2
        \\&=
        u_j \left(1 - \frac{2^{k+1} + 1}{k'} \right) - 2 t_{j-1} - 2
        \\&\geq
        \frac{u_j}{2} - 2 t_{j-1} & \text{def of $k'$}
        \\&=
        \frac{1}{4} u_j + t_{j-1}
        \\&\geq
        \frac{1}{4} (|S_j| + |S_1 \ldots S_{j-1}|) & \text{def of $u_j$ and $t_{j-1}$}
        \\&=
        \frac{1}{4} |S_1 \ldots S_{j}|.
    \end{align*}
    Because $\phi(j)$ revisits every pair $(k,k')$, with $k' = 2^{2^{k+1}}$, for every $k$, there exists $\widehat{k}=k'+c$ such that, on infinitely many $j$, the above inequality holds. Hence $S$ is finite-state deep.
}

\opt{journal}{\begin{proof}\proofThmExistsFSD\end{proof}}


\section{Polynomial-Time Depth}
Because the time bound defining polynomial-time depth is in terms of the characteristic sequence of a language, we focus on the class $\E$ of languages decidable in time $2^{c|s_n|}$ for a fixed $c\in\N$, or equivalently, $n^c$, where $n$ is the length of the characteristic sequence of a language up to the string $s_n$.
\subsection{Measure in $\E$}
    We use Lutz's measure theory for the complexity class $\E$, which we now briefly describe. See \cite{Lutz:QSET} for more details.

    Measure on $\E$ is obtained by imposing appropriate resource bounds on a game theoretical characterization of the classical Lebesgue measure of subsets of $\C$. A \emph{martingale} is a function $d:\binary\to [0,\infty)$ such that, for every $w \in \binary$,
    $$d(w) = \frac{d(w0) + d(w1)}{2}.$$
    We say that a martingale $d$ \emph{succeeds} on a language $L$ if  $\limsup_{n\to\infty} d(L \pre n) = \infty$. Intuitively, $d$ is a gambler that bets money on each successive bit of $\chi_L$, doubling the money bet on the bit that occurs, and losing the rest. It succeeds by making unbounded money.

    A class of languages $\calC$ has \emph{$\p$-measure zero}, and we write $\mup(\calC)=0$, if there is a polynomial-time computable martingale that succeeds on every language in $\calC$. $\calC$ has \emph{measure zero in $\E$}, denoted $\muE \calC = 0$, if $\calC\cap \E$ has $\p$-measure zero. A class $\calC$ has \emph{$\p$-measure one}, denoted $\mu_\p(\calC)=1$, if $\overline{\calC}$ has $\p$-measure zero, where $\overline{\calC}$ denotes the complement of $\calC$, and $\calC$ has \emph{measure one in $\E$}, denoted $\muE \calC = 1$, if $\E - \calC$ has $\p$-measure zero.
    We say that a language $L$ is \emph{$\E$-random} if the singleton $\{L\}$ does not have $\p$-measure zero.

    Measure in $\E$ yields a size notion on the class $\E$ similar to Lebesgue measure on the Cantor space. Subsets of $\E$ that have $\p$-measure zero are then ``small subsets of $\E$''; for example, the singleton set $\{ L \}$ for any $L\in\E$. $\E$, being the largest subset of itself, has $\p$-measure one.

\subsection{Polynomial-Time Depth}
    This section proposes a variation of depth based on polynomial-time oblivious predictors, which, given a language $L$, try to predict $L[n]$ (i.e., the membership of $s_n$ in $L$), \emph{without} having access to $L[0 \twodots n-1]$. This is in contrast to a martingale, where the bet on $L[n]$ is by definition a function of $L[0 \twodots n-1]$. Intuitively, $L$ is polynomial-time deep if giving a polynomial-time predictor more time allows it to predict bits of $L$ with significantly greater accuracy.

    An \emph{oblivious predictor} is a function $P:\bool{*}\times \bool{} \rightarrow [0,1]$ such that, for all $x\in\binary$, $P(x,0)+P(x,1)=1$. Intuitively, when trying to predict a language $L$, $P(x,1)$ is the probability with which the predictor predicts that $x\in L$. To measure how well a predictor $P$ predicts $L$, we consider its associated martingale $p:\bool{*}\rightarrow [0,\infty)$ given by
    $$p(L\harp n)= 2^{n}\prod_{y \leq s_n} P(y,L(y)).$$
    We shall consider predictors $P$ such that $P(s_n,b)$ is computable in time polynomial in $n$ (hence computable in time $2^{c |s_n|}$ for some constant $c$), and call such a $P$ a \emph{polynomial-time oblivious predictor}, and we call the martingale $p$ its \emph{polynomial-time oblivious martingale (pom)}, with the convention that predictors are given in uppercase and pom in lowercase.

    \begin{defn}
    A language $L$ is \emph{polynomial-time deep} if there exists $a>0$ such that, for all pom $p$, there exists a pom $p'$ such that, for infinitely many $n\in\N$,
    $$\frac{p'(L\harp n)}{p(L\harp n)} \geq a\log n,$$
    with the convention that $\frac{1}{0}=\infty$.
    $L$ is \emph{polynomial-time shallow} if it is not polynomial-time deep.
    \end{defn}

     Languages that are trivial or random for $\E$ are polynomial-time shallow.

    \begin{prop} \label{prop-easy-random-not-poly-deep}
    Let $L$ be a language.
        \begin{enumerate}
            \item   If $L\in\E$, then $L$ is polynomial-time shallow.
            \item   If $L$ is $\E$-random, then $L$ is  polynomial-time shallow.
        \end{enumerate}
    \end{prop}

\newcommand{\proofPropEasyRandomNotPolyDeep}{
    For the first item, let $a>0$ and $L\in\E$. Then there exists a pom $p$ that predicts $L$ correctly on every string, i.e., $p(L\harp n)=2^n$ for every $n\in\N$. Hence for any pom $p'$ we have
    $$\frac{p'(L\harp n)}{p(L\harp n)} \leq \frac{2^n}{2^n} = 1 < a\log n$$
    for all but finitely many $n$. Because $a$ is arbitrary, $L$ is  polynomial-time shallow.

    For the second item let $a>0$ and $L$ be $\E$-random. Then for any pom $p$ there exists $c_p\in\N$ such that for every $n\in\N$, $p(L\harp n) < c_p$. Fix a pom $p$ such that $p(L \pre n) \geq 1$ for all $n\in\N$. Then for any pom $p'$ we have
    $$\frac{p'(L\harp n)}{p(L\harp n)} \leq \frac{c_{p'}}{1} < a\log n$$
    for all but finitely many $n$. Thus $L$ is polynomial-time shallow.
}
\opt{journal}{\begin{proof}\proofPropEasyRandomNotPolyDeep\end{proof}}

\newcommand{\Ereduc}{$\E$-Lb-M }
\newcommand{\leqm}[1]{\leq_{\mathrm{M}}^{#1}}
\newcommand{\geqm}[1]{\geq_{\mathrm{M}}^{#1}}
\newcommand{\leqE}{\leqm{\E,\mathrm{Lb}}}

\subsection{Slow Growth Law}
    Let $f:\binary\to\binary$.
    We say $f$ is \emph{monotone} if, for all $x,y\in\binary$, $x < y \implies f(x) < f(y)$.
    Given $l:\N\to\N$, we say $f$ is \emph{$l$-bounded} if, for all $x\in\binary$, $|f(x)| \leq l(|x|)$.
    Given two languages $L_1,L_2$ and a time bound $t:\N\to\N$ and length bound $l:\N\to\N$, we say that $L_1$ is \emph{$t$-time $l$-bounded monotone many-one reducible} to $L_2$ (abbreviated \emph{$t$-$l$-M reducible}), and we write $L_1 \leqm{t,l} L_2$, if there is a Turing machine $M$ computing a monotone, $l$-bounded reduction $f:\binary\to\binary$ such that, on input $s_n$, $M$ halts in at most $t(|s_n|) = t(\log n)$ steps and outputs $f(s_n)\in\binary$ such that $s_n \in L_1$ if and only if $f(s_n) \in L_2$.
    We say $L_1$ is \emph{$\E$-time linearly bounded monotone many-one reducible} to $L_2$ (abbreviated \emph{\Ereduc reducible}), and we write $L_1 \leqE L_2$, if there exists $c\in\N$ such that $L_1 \leqm{2^{c|s_n|},c|s_n|} L_2$.
    We follow the convention of letting $n$ refer to the length of a characteristic sequence, rather than the length of the input string $s_n$. Therefore, equivalently, $L_1 \leqm{n^c,n^c} L_2$; i.e., $f(s_n)$ is computable in time $n^c$, and, if $m\in\N$ is such that $s_m=f(s_n)$, then $m \leq n^c$.

    The following result shows that shallow sequences cannot be transformed into deep ones by simple processes.

    \begin{theorem}[Polynomial-time slow growth law]\label{t.sgl}
        Let $L_1,L_2$ be languages such that $L_1 \leqE L_2$. If $L_1$ is  polynomial-time deep, then $L_2$ is polynomial-time deep.
    \end{theorem}

\newcommand{\proofThmPolySGL}{
    Let $f:\binary\to\binary$ be the \Ereduc reduction from $L_1$ to $L_2$, and let $c\in\N$ such that $f$ is computable in time $n^c$ ($= 2^{c |s_n|}$), so that, for all $x\in\binary$, $x \in L_1 \iff f(x) \in L_2$, and $|f(x)| \leq c |x|$.

    Let $p_2$ be any pom, such that $P_2$ is computable in time $n^k$ for some $k$. Consider the pom $p_1$, where, for all $x\in\binary$ and $b\in\{0,1\}$, $P_1(x,b)=P_2(f(x),b)$. Then, if $x=s_n$, $P_1$ is computable in time $n^c + |f(x)|^k \leq n^c + n^{ck}$, so $P_1$ is computable in time polynomial in $n$. Since $L_1$ is polynomial-time deep, there exist a pom $p_1'$, a constant $a>0$, and an infinite set $N \subseteq \N$ such that, for every $n\in N$,

    \begin{equation} \label{eq-L1deep}
    \frac{p_1'(L_1\harp n)}{p_1(L_1\harp n)} \geq a \log n.
    \end{equation}

    Consider the following pom $p_2'$, where, for all $y\in\binary$ and $b\in\{0,1\}$,
    \[
        P'_2(y,b)=
        \begin{cases}
            P'_1(f^{-1}(y),b)& \text{if $f^{-1}(y)$ exists},\\
            P_2(y,b)& \text{otherwise.}
        \end{cases}
    \]
    For all $n\in\N$, define $m_n\in\N$ such that $s_{m_n} \triangleq f(s_n)$. Because $f$ is monotone, it is 1-1. Thus, if $f^{-1}(y)$ exists, then it is unique. Because $f$ is monotone, $n \leq m_n$. Therefore, letting $y=s_{m_n}$ and $x=f^{-1}(y)=s_n$, $x$ can be computed from $y$ in time polynomial in $m_n$ by searching all strings $w \leq s_{m_n}$ and checking whether $f(w)=y$, which takes at most $m_n n^c \leq m_n^{c+1}$ steps. Hence $P'_2$ is polynomial-time computable.

    For every $n\in N$,
    \begin{align*}
        a \log n
        &\leq \frac{p_1'(L_1\harp n)}{p_1(L_1\harp n)} \qquad &\text{by \eqref{eq-L1deep}}\\
        &= \frac{\prod_{x\leq s_n}P_1'(x,L_1(x))}{\prod_{x\leq s_n}P_1(x,L_1(x))}\\
        &= \frac{\prod_{x\leq s_n}P_2'(f(x),L_2(f(x)))}{\prod_{x\leq s_n}P_2(f(x),L_2(f(x)))}\\
        &= \frac{\prod_{y\in f(\{s_0,\ldots, s_n\})}P_2'(y,L_2(y))}{\prod_{y\in f(\{s_0,\ldots, s_n\})}P_2(y,L_2(y))} \qquad &\text{putting $y=f(x)$}\\
        &= \frac{\prod_{y\leq f(s_n)}P_2'(y,L_2(y))}{\prod_{y\leq f(s_n)}P_2(y,L_2(y))}
        \qquad & {P'_2(y,b)=P_2(y,b) \text{ if } f^{-1}(y)  \atop \text{ undefined, and } f \text{ monotone} } \\
        &= \frac{p_2'(L_2\harp m_n)}{p_2(L_2\harp m_n)}.
    \end{align*}
    Because $f$ is linearly bounded, for all $n\in\N$, $m_n \leq n^c$. Thus for any $n\in N$,
    \[
        \frac{a}{c} \log m_n
        \leq
        \frac{a}{c} \log n^c
        =
        a \log n
        \leq
        \frac{p_2'(L_2\harp m_n)}{p_2(L_2\harp m_n)}.
    \]
    Thus the constant $a / c$ testifies that $L_2$ is polynomial-time deep.
%
}

\opt{journal}{\begin{proof}\proofThmPolySGL\end{proof}}

\subsection{Languages that are Useful for $\E$}
    In \cite{Benn88} Bennett showed that the halting language is deep, and Juedes, Lathrop, and Lutz \cite{Lutz:CDR} generalized this result by showing every weakly useful \cite{Lutz:CDR,Lutz:WUS} language is deep. We prove a polynomial-time version of the result of Juedes, Lathrop, and Lutz, namely, that every \Ereduc weakly useful language is polynomial-time deep.

    Following the definition of weakly useful languages from \cite{Lutz:CDR} and \cite{Lutz:WUS}, we define a language $L$ to be \Ereduc weakly useful if the set of languages in $\E$ that are reducible to $L$ -- within a fixed time and length bound -- is not small (does not have measure zero in $\E$). Intuitively, an $\E$-useful language is somewhere in between an $\E$-hard language and a trivial language, in the sense that the language does not necessarily enable one to decide \emph{all} languages in $\E$, but rather a nonnegligible subset of them. Note, however, that an $\E$-hard (for instance, under polynomial-time many-one reductions) language may not necessarily be \Ereduc weakly useful because of the requirements that an \Ereduc reduction be monotone and linearly bounded.

    \begin{defn}
    A language $L$ is \emph{\Ereduc weakly useful} if there is a $c\in\N$ such that the set of languages $2^{c |s_n|}$-$c|s_n|$-M reducible to $L$ does not have measure zero in $\E$, i.e., if
    $$\muE{L^{\geqm{2^{c |s_n|},c|s_n|}}} \neq 0$$
    where
    $$L^{\geqm{2^{c |s_n|},c|s_n|}} = \setr{A}{A \leqm{2^{c |s_n|},c|s_n|} L}.$$
    \end{defn}

    In other words, a language $L$ is weakly useful if a nonneglible subset of $\E$ monotonically many-one reduces to $L$ within a fixed exponential time bound and fixed linear length bound.
    An example of an \Ereduc weakly useful language is the halting language for $\E$, defined as follows. Fix a standard linear-time computable invertible encoding of pairs of strings $(x,y)\mapsto \langle x,y\rangle$. Let $M_1,M_2,\ldots$ be an enumeration of machines deciding languages in $\E$, where machine $M_i$ runs in time $2^{i|s_n|}$. The $\E$-halting language is given by $H_{\E} = \setl{ \langle 0^i,x \rangle }{ M_i \text{ accepts } x }$. It is easy to verify that access to the $\E$-halting language allows one to decide every language $L_i\in\E$, decided by machine $M_i$, using the $1.01|s_n|$-time-bounded, $1.01|s_n|$-length-bounded, monotone reduction $f(x) = \langle 0^i,x \rangle$; i.e., $\E \subseteq H_{\E}^{\geqm{1.01|s_n|,1.01|s_n|}},$ whence $H_{\E}$ is \Ereduc weakly useful.

    For every $g:\N\to\N$ and pom $p$ define
    $$D^g_p = \setr{ L\in\C } { (\exists \text{ pom } p') (\existsio n\in\N)\ \frac{p'(L\harp n)}{p(L\harp n)} \geq g(n) }.$$
    Note that $L$ is polynomial-time deep if and only if there exists $a>0$ such that, for all pom $p$, $L \in D^{a \log n}_p$.

\opt{journal}{The next lemma shows that most languages in $\E$ are contained in $D^g_p$ for fixed $g$ and $p$.}

    \begin{lem} \label{lem-D-measure-1}
        For any $g:\N\to\N$ such that $g(n) = o(2^n)$ and any pom $p$, $\muE{D^g_p} = 1$.
    \end{lem}

\newcommand{\proofLemDMeasureOne}{
     Let $g,p$ be as in the statement of the lemma. Let $L \in \E - D^g_p$. It suffices to show that $p$ succeeds on $L$. $L\in\E$ implies the existence of a pom $p'$ such that for any $n\in\N$, $p'(L\harp n)=2^{n}$. $L \not\in D^g_p$ implies that for all pom $p'$ there are infinitely many $n\in\N$ such that $p(L\harp n)> p'(L\harp n) / g(n)$. Thus $p(L\harp n)> 2^n / g(n)$, which grows unboundedly as $n\to\infty$ because $g=o(2^n)$; i.e., $p$ succeeds on $L$.
}

\opt{journal}{\begin{proof}\proofLemDMeasureOne\end{proof}}

    \begin{theorem}\label{t.wuisdeep}
        Every \Ereduc weakly useful language is polynomial-time deep.
    \end{theorem}
\newcommand{\proofThmWeakUsefIsDeep}{
    Let $B$ be any \Ereduc weakly useful language, i.e. $\muE{B^{\geqm{2^{c |s_n|},c|s_n|}}} \neq 0$ for some $c\in\N$. Let $a=\frac{1}{c}$ and let $p_2$ be any pom. It suffices to show that $B\in D^{a \log n}_{p_2}$. Let $p_1$ be constructed from $p_2$ as in the proof of Theorem \ref{t.sgl}. By Lemma \ref{lem-D-measure-1}, $\muE{D^{\log n}_{p_1}} = 1$. Thus $D^{\log n}_{p_1} \cap B^{\geqm{2^{c |s_n|},c|s_n|}} \neq \emptyset$, whence there exists a language $A \in D^{\log n}_{p_1} \cap B^{\geqm{2^{c |s_n|},c|s_n|}}$. Thus $A \leqm{2^{c |s_n|},c|s_n|} B$ and $A\in D^{\log n}_{p_1}$, so by the proof of Theorem \ref{t.sgl}, $B\in D^{a \log n}_{p_2}$.
}

\opt{journal}{\begin{proof}\proofThmWeakUsefIsDeep\end{proof}}

    \begin{cor} \label{cor-H_E-deep}
        $H_\E$ is polynomial-time deep.
    \end{cor}

    \begin{cor} \label{cor-E-trivial-not-useful}
    No language in $\E$ is \Ereduc weakly useful.
    \end{cor}

    \begin{cor} \label{cor-E-random-not-useful}
    No $\E$-random language is \Ereduc weakly useful.
    \end{cor}

    No decidable language is deep in the sense of Bennett \cite{Benn88} (see also \cite[Corollary 5.7]{Lutz:CDR}). However, the halting language $H$ is deep and, while not decidable, is computably enumerable. Compare this with the fact that Corollary \ref{cor-E-trivial-not-useful} (or a simple diagonalization) implies that $H_\E \not\in \E$. It is easy to verify, however, $H_\E \in \DTIME(2^{|s_n|^2}) \subseteq \EXP$. Thus, polynomial-time depth mirrors Bennett's depth in that $\E$-decidable languages are not polynomial-time deep, but polynomial-time deep languages can be found ``close'' to $\E$. Similarly, Lemma \ref{lem-D-measure-1} tells us, in an analogous fashion to Corollary 5.10 of \cite{Lutz:CDR}, that ``partially deep'' sequences can be found in abundance in $\E$.


\begin{ack}
We thank Jim Lathrop for many useful and stimulating discussions in the early stages of this research.
\end{ack}


\bibliographystyle{abbrv}
\bibliography{dim,rbm,main,random,dimrelated}

\begin{thebibliography}{10}

\bibitem{AFVV06}
L.~Antunes, L.~Fortnow, D.~van Melkebeek, and N.~Vinodchandran.
\newblock Computational depth: Concept and applications.
\newblock {\em Theoretical Computer Science}, 354(3):391--404, 2006.
\newblock Special issue for selected papers from the 14th International
  Symposium on Fundamentals of Computation Theory.

\bibitem{Athreya:ESDAICC}
K.~B. Athreya, J.~M. Hitchcock, J.~H. Lutz, and E.~Mayordomo.
\newblock Effective strong dimension, algorithmic information, and
  computational complexity.
\newblock {\em SIAM Journal on Computing}.
\newblock To appear. Preliminary version appeared in {\em V. Diekert and M.
  Habib (eds.), Proceedings of the 21st International Symposium on Theoretical
  Aspects of Computer Science}, Springer Lecture Notes in Computer Science,
  Montpellier, France, March 25-27, 2004, pp. 632--643.

\bibitem{Benn88}
C.~H. Bennett.
\newblock Logical depth and physical complexity.
\newblock In R.~Herken, editor, {\em The Universal Turing Machine: A
  Half-Century Survey}, pages 227--257. Oxford University Press, London, 1988.

\bibitem{Bore09}
E.~Borel.
\newblock Sur les probabilit\'es d\'enombrables et leurs applications
  arithm\'etiques.
\newblock {\em Rendiconti del Circolo Matematico di Palermo}, 27:247--271,
  1909.

\bibitem{Bourke:ERFSD}
C.~Bourke, J.~M. Hitchcock, and N.~V. Vinodchandran.
\newblock Entropy rates and finite-state dimension.
\newblock {\em Theoretical Computer Science}, 349:392--406, 2005.
\newblock To appear.

\bibitem{Dai:FSD}
J.~J. Dai, J.~I. Lathrop, J.~H. Lutz, and E.~Mayordomo.
\newblock Finite-state dimension.
\newblock {\em Theoretical Computer Science}, 310:1--33, 2004.

\bibitem{DotMos06}
D.~Doty and P.~Moser.
\newblock Finite-state dimension and lossy decompressors.
\newblock Technical Report cs.CC/0609096, Computing Research Repository, 2006.

\bibitem{Lutz:WUS}
S.~A. Fenner, J.~H. Lutz, E.~Mayordomo, and P.~Reardon.
\newblock Weakly useful sequences.
\newblock {\em Information and Computation}, 197:41--54, 2005.

\bibitem{Hitchcock:FDLLU}
J.~M. Hitchcock.
\newblock Fractal dimension and logarithmic loss unpredictability.
\newblock {\em Theoretical Computer Science}, 304(1--3):431--441, 2003.

\bibitem{Huff59a}
D.~A. Huffman.
\newblock Canonical forms for information-lossless finite-state logical
  machines.
\newblock {\em IRE Trans. Circuit Theory CT-6 (Special Supplement)}, pages
  41--59, 1959.
\newblock Also available in E.F. Moore (ed.), Sequential Machine: Selected
  Papers, Addison-Wesley, 1964, pages 866-871.

\bibitem{Lutz:CDR}
D.~W. Juedes, J.~I. Lathrop, and J.~H. Lutz.
\newblock Computational depth and reducibility.
\newblock {\em Theoretical Computer Science}, 132(1--2):37--70, 1994.

\bibitem{Koha78}
Z.~Kohavi.
\newblock {\em Switching and Finite Automata Theory (Second Edition)}.
\newblock McGraw-Hill, 1978.

\bibitem{Lutz:RCD}
J.~I. Lathrop and J.~H. Lutz.
\newblock Recursive computational depth.
\newblock {\em Information and Computation}, 153(2):139--172, 1999.

\bibitem{LiVi97}
M.~Li and P.~M.~B. Vit\'{a}nyi.
\newblock {\em An Introduction to Kolmogorov Complexity and its Applications}.
\newblock Springer-Verlag, Berlin, 1997.
\newblock Second Edition.

\bibitem{Lutz:AEHNC}
J.~H. Lutz.
\newblock Almost everywhere high nonuniform complexity.
\newblock {\em J.~Comput. Syst. Sci.}, 44(2):220--258, 1992.

\bibitem{Lutz:QSET}
J.~H. Lutz.
\newblock The quantitative structure of exponential time.
\newblock In L.~A. Hemaspaandra and A.~L. Selman, editors, {\em Complexity
  Theory Retrospective {II}}, pages 225--254. Springer-Verlag, 1997.

\bibitem{SchSti72}
C.~P. Schnorr and H.~Stimm.
\newblock Endliche {Automaten} und {Zufallsfolgen}.
\newblock {\em Acta Informatica}, 1:345--359, 1972.

\bibitem{Shannon48}
C.~E. Shannon.
\newblock A mathematical theory of communication.
\newblock {\em Bell System Technical Journal}, 27:379--423, 623--656, 1948.

\bibitem{SheLemZiv95}
D.~Sheinwald, A.~Lempel, and J.~Ziv.
\newblock On encoding and decoding with two-way head machines.
\newblock {\em Information and Computation}, 116(1):128--133, Jan. 1995.

\bibitem{ZivLem78}
J.~Ziv and A.~Lempel.
\newblock Compression of individual sequences via variable-rate coding.
\newblock {\em IEEE Transaction on Information Theory}, 24:530--536, 1978.

\end{thebibliography}

\opt{cwa}{
\newpage
\section{Appendix}
This section contains proofs of some results from the main text, as well as auxiliary results needs for these proofs.

\lemmaOneAndTwo



\begin{proof}[of Proposition \ref{propFSTrivRandShallow}]
\proofPropFSTrivRandShallow
\qedllncs\end{proof}

\begin{proof}[of Theorem \ref{thm-sgl}]
\proofThmSGL
\qedllncs\end{proof}

\thmExistsFSDSetup
\begin{proof}[of Theorem \ref{thm-exists-fsd}]
\proofThmExistsFSD
\qedllncs\end{proof}

\begin{proof}[of Proposition \ref{prop-easy-random-not-poly-deep}]
\proofPropEasyRandomNotPolyDeep
\qedllncs\end{proof}

\begin{proof}[of Theorem \ref{t.sgl}]
\proofThmPolySGL
\qedllncs\end{proof}

\begin{proof}[of Theorem \ref{t.wuisdeep}]
\proofThmWeakUsefIsDeep
\qedllncs\end{proof}

\begin{proof}[of Lemma \ref{lem-D-measure-1}]
\proofLemDMeasureOne
\qedllncs\end{proof}

}

\end{document}